# Rashba-splitting-induced topological flat band detected by anomalous resistance oscillations beyond the quantum limit in ZrTe$_5$

Dong Xing[1,2], Bingbing Tong[1], Senyang Pan[3], Zezhi Wang[1,2], Jianlin Luo[1,2], Jinglei Zhang[3], and Cheng-Long Zhang[1*]

[1]*Beijing National Laboratory for Condensed Matter Physics, Institute of Physics, Chinese Academy of Sciences, Beijing 100190, China*

[2]*School of Physical Sciences, University of Chinese Academy of Sciences, Beijing 100049, China.*

[3] *High Magnetic Field Laboratory, HFIPS, Chinese Academy of Sciences, Hefei 230031, China*

Corresponding author: chenglong.zhang@iphy.ac.cn

**Topological flat band, on which the kinetic energy of topological electrons is quenched, represents a platform for investigating the topological properties of correlated systems. Recent experimental studies on flattened electronic bands have mainly concentrated on 2-dimensional materials created by van der Waals heterostructure-based engineering. Here, we report the observation of a topological flat band formed by polar-distortion-assisted Rashba splitting in a 3-dimensional Dirac material $ZrTe_5$. The polar distortion and resulting Rashba splitting on the band are directly detected by torque magnetometry and the anomalous Hall effect, respectively. The local symmetry breaking further flattens the band, on which we observe resistance oscillations beyond the quantum limit. These oscillations follow the temperature dependence of the Lifshitz–Kosevich formula but are evenly distributed in $B$ instead of $1/B$ in high magnetic fields. Furthermore, the cyclotron mass anomalously gets enhanced about $10^2$ times at field ~20 T. These anomalous properties of oscillations originate from a topological flat band with quenched kinetic energy. The topological flat band, realized by polar-distortion-assisted Rashba splitting in the 3-dimensional Dirac system $ZrTe_5$, signifies an intrinsic platform without invoking moiré or order-stacking engineering, and also opens the door for studying topologically correlated phenomena beyond the dimensionality of two.**

Flat electronic bands harbor exotic quantum behaviors due to the quenched kinetic energy and subsequently dominated Coulomb interaction. The fractional quantum Hall effect[1,2] is an archetypical 2-dimensional (2D) flat band system. Recently developed moiré-engineered 2D twisted bilayer graphene[3-5] and multilayer graphene in certain stacking order[6-8] are other examples of realizing topological flat bands. Flat bands are also theoretically predicted in some stoichiometric 3-dimensional (3D) materials forced by certain geometric lattices, like the Kagome or Lieb lattices[9,10]. Flat-band-induced correlation in 3D topological systems is crucial for realizing correlated 3D topological effects, like correlation on Weyl semimetals (WSM) and possible axionic dynamics[11-13]. Despite theoretical advancement in establishing the material database of topological flat bands[14], experimental realization of an isolated topological flat band around Fermi level (FL) in 3D stoichiometric materials remains elusive.

In the twisted bilayer graphene, the two preconditions for realizing a 2D topological flat band are: 1. the pristine Dirac material graphene; 2. moiré superlattice as the method for flattening the band. We now ask a question: can we find a counterpart in 3D? If there is, then which material is the 3D counterpart of graphene? How do we flatten the energy band or enlarge the unit cell in 3D? Here, we report that $ZrTe_5$[15], as a typical Dirac material in 3D, can meet the first precondition; the polar distortion at low temperatures in $ZrTe_5$ meets the second precondition without invoking van der Waals heterostructure-based engineering. The polar-distortion-assisted Rashba splitting in $ZrTe_5$ is evidenced by torque magnetometry and the anomalous Hall effect (AHE), which shows the existence of a topological flat band in the magnetic field, on which anomalous resistance oscillations appear beyond the quantum limit. The cyclotron mass is enhanced by an order of $10^2$, consistent with the topological flat band. Our work also highlights the importance of local symmetry breaking in topological materials.

$ZrTe_5$ was initially proposed as a typical candidate for a quantum spin Hall insulator in 2D limit[15]. As shown in Fig. 1**a**, 2D $ZrTe_5$ layers stack along ***b*** axis (here, we use ***a, b, c*** and -*x, z, y* interchangeably) with a $ZrTe_3$ chain that runs along ***a*** axis. The topological property of 3D ZrTe$_5$ is sensitively dependent on the inter- and intra-layer coupling strengths. ZrTe$_5$ is located near the boundary of weak topological (WTI) and strong topological insulators (STI) as typical Dirac material[15], promoting many exotic phenomena[16-22]. Therefore, the properties of ZrTe$_5$ are sensitively dependent on crystal growth methods, namely the chemical vapor transfer (CVT) and flux processes. Samples grown from Te-flux are more stoichiometric and closer to the phase boundary[23-26]. As shown in Fig. 1**b**, temperature-dependent resistivity measured on flux-grown single crystals exhibits semiconducting-like behavior, a typical profile in narrow-gapped semiconductors[27,28]. We focus on electrical transport within ***ac*** plane with current ***i*** along ***a*** axis as illustrated in the inset of Fig. 1**b**. Hall measurements (Fig. 1**c**) show hole is the only carrier down to 2 K. Hall conductivity $\sigma_{xy}$ (Fig. 1**d**) fitted by Drude model $\sigma_{xy}^{Drude} = \frac{pe\mu^2 B}{1+\mu^2 B^2}$, where $p$ is the carrier density and $\mu$ is the mobility, shows the existence of an anomalous term $\sigma_{xy}^A$, as indicated by the pink shadowed area. As shown in Fig. 1**e**, $p$ is ultralow ~ $3\times10^{14}$ cm$^{-3}$ and $\mu$ is as high as $10^6$ cm$^2$ V$^{-1}$s$^{-1}$ at low temperatures, crucial for realizing the topological flat band and anomalous resistance oscillations. Therefore, ZrTe$_5$ sample synthesized in this work is a 3D counterpart of graphene.

We then investigate the possible local modifications on the Dirac band, which might provide the clue for forming a topological flat band. In these flux-grown samples, preliminary evidence of polar distortion is reported by nonlinear transport[26], while direct evidence is still lacking. We adopted torque magnetometry to measure the magnetic susceptibility tensor $\chi_{ij}$ defined by $M_i = \chi_{ij} H_j$, where $M_i$ is the magnetization. $\chi_{ij}$ directly reflects the underlying point group symmetries due to Neumann's principle. Magnetic torque is defined as $\tau = \mu_0 V M \times H$, where $\mu_0$ is the vacuum permeability, and $V$ is the volume of the sample. The space group of ZrTe$_5$ is $Cmcm$ (No. 63) with a point group $D_{2h}$, under which the only permitted tensor elements in $\chi_{ij}$ are $\chi_{aa}, \chi_{cc}$ and $\chi_{bb}$. In our torque setup (inset of Fig. 2**a**), the cantilever picks up $\tau_{a(x)} = \tau_{2\theta} = A_1 sin2\theta$, where $A_1 = \frac{1}{2}\mu_0 V H^2 (\chi_{cc} - \chi_{bb})$. Therefore, we anticipate a pure $sin2\theta$ relation in angle-dependent $\tau_a$. As shown in Fig. 2**a**, $\tau_a$ at 7 T globally shows $\pi$ periodicity consistent with $sin2\theta$ relation. However, the negative and positive amplitudes, noted as Amp+ and Amp- in Fig. 2**a**, show asymmetry against a pure $sin2\theta$ relation. Further, we find the torque signal at 4 K can be well-fitted by $\tau_a = \tau_{2\theta} + A_2 \sin^2\theta$. The appearance of $A_2$ term directly shows the original orthorhombic symmetry is broken at low temperature, which is consistent with polarity (***P***) along the out-of-plane ***b***-axis (***P**//**b***) in our nonlinear transport results (see Fig. S2 and SI section 2 for details of symmetry analyses). We now show that the AHE is the direct consequence of this polar distortion. As shown in Fig. 2**b**, $\sigma_{xy}^A$ starts to show up and exhibits plateau-like structures beyond a critical magnetic field $B_p$. As shown in Fig. 2**c**, we extract the temperature-dependent coefficients $A_1$, $A_2$, and $\sigma_{xy}^A$, and plot the ratios of $A_{1,2}(T)/A_{1,2}(270\ K)$ and $\sigma_{xy}^A$ together. $A_1(T)/A_1(270\ K)$ exhibits moderate temperature dependence and dominates over the total torque $\tau_a$ in the whole temperature range (see Fig. S1**a** for the raw data). However, $A_2(T)$ is almost negligible at temperatures higher than 150 K, below which the ratio of $A_2(T)/A_2(270\ K)$ suddenly gets enhanced and reaches a value of ~30 at 2 K, indicating the emergence of polar distortion at 150 K. Furthermore, as shown in Fig. 2**c**, $\sigma_{xy}^A$ shows up around 150 K, where the $A_2$ term suddenly gets enhanced. The concurrence of the polar distortion and $\sigma_{xy}^A$ indicates that the AHE is locked to the polar-distortion-induced band modifications, which is also consistent with the fact that no AHE is observed in CVT samples[20].

ZrTe$_5$ is a nonmagnetic material with time-reversal symmetry. Nevertheless, the observed AHE takes a profile like the AHE of ferromagnets with saturating plateaus. Based on our three observations: ***1***. The existence of polar distortion ***P**//**b***; ***2***. The coincidence of the onsets of the AHE and polar distortions; ***3***. The absence of in-plane Hall effect in our samples (see Fig. S3 and SI section 3 for the in-plane Hall discussions). Considering the quasi-2D nature of ZrTe$_5$, we interpret this behavior by invoking a Rashba model-based mechanism usually adopted for explaining the intrinsic AHE in magnetic materials[29]. With polar distortion ***P*** along ***b*** axis, the typical Rashba model appears as $H = \boldsymbol{\alpha} \cdot (\boldsymbol{\sigma} \times \boldsymbol{k}_\parallel) + \Delta(B)\sigma^z$ with the time-reversal breaking term $\Delta(B)$, here momentum $\boldsymbol{k}_\parallel$ in the ***ac*** plane, $\boldsymbol{\alpha}$ is the strength of Rashba splitting, and $\boldsymbol{\sigma}$ is the Pauli matrix for real spins. Furthermore, we do not include $k_z$ dispersion in the Rashba model due to the parabolic relation along $k_z$[30]. As shown in Fig. 2**d**, the band splitting is very weak, and the crossing point is near the band edges due to tiny polar distortions, which means only samples with ultralow carrier density can access this region. When an out-of-plane magnetic field ($B_\perp$) is applied, the crossing point on the Rashba bands is gapped out due to a significant Zeeman effect $\Delta(B)$, the region near the gap becomes Berry curvature hot spots. With increasing magnetic field, the FL located near the band edge falls into the Rashba gap at $B_p$, results in a saturated $\sigma_{xy}^A$ in higher fields[29]. When we tilt the direction of a magnetic field to the *ac* plane, no in-plane Hall signal is detected (Fig. S3**a** & 3**b**), which is consistent with the Rashba model, because the in-plane magnetic field can only shift the Rashba crossing without opening a gap, then no AHE is expected to appear.

The prominent feature of the formed Rashba band is that the extremum of the gapped band gets flatter with increasing magnetic field, creating an ideal, isolated topological flat band around FL, different from the proposal of NLSM-based flat band used to explain the results of nonlinear transport[31]. The origin of tiny polar distortions in the flux-grown $ZrTe_5$ sample is not yet clear because no structural transition is observed by x-ray diffraction down to 10 K[26], which might also indicate this weak polarity $\boldsymbol{P}$ is unable to drive the parent TIs to WSM phase via the Murakami's scheme[32]. As we know, the possible defects and disorder-induced polar distortion, as essential roles of local symmetry breakings, will lift the valley or spin degeneracies, and form a supercell[33-35]. As the main result of this work, we propose in this work that the effect of polar distortion together with Rashba splitting on the band edge, by forming the supercell, is similar to moiré engineering. As shown in the right column of Fig. 2**d**, the local symmetry breaking enlarges the unit cell, induces multiple Rashba splittings, and creates a topological flat band.

As we will show, the flatness on the modified band is further supported by the observation of anomalous resistance oscillations beyond the quantum limit. The carrier density of sample S75 is $3\times10^{14}$ cm$^{-3}$ at 2 K, corresponding to a quantum limit less than 0.05 T ($B \parallel b$ axis), beyond which there are no quantum oscillations. However, as shown in Fig. 3**a** and 3**b**, we observe strong resistance oscillations in sample S75, and reproduced in another sample S74, where similar behaviors persist up to 18 T (Fig. 3**c**). As noted by blue vertical lines in Fig. 3**c**, we find evenly-distributed oscillations at fields higher than 3 T, which is clearly exhibited in $\Delta\sigma_{xx}$ (inset of Fig. 3**c**). This linear-in-$B$ relation is against the normal Shubnikov-de Haas (SdH) oscillations evenly distributed in $1/B$ and also different from the logarithmic oscillations[36]. As shown in Fig. 3**d**, we index the oscillations by integers, which is found to be well fitted by $n \sim \frac{C_0}{B} + C_1 * B$, where the $C_0$ term represents contribution from normal $1/B$, and $C_1$ term represents the additional contribution from linear-in-$B$. The same fitting is also employed to fit fan diagrams obtained from $\Delta\sigma_{xx}$ (Fig. 3**e** & 3**f**), where linear-in-$B$ trend is highlighted by shadowed area. As shown in Fig. 3**g** & 3**h**, we also tilt the magnetic field in **ba** and **bc** planes (defined as $\theta, \phi$), and find the oscillations gradually shift towards high magnetic fields. As summarized in Fig. 3**i** & 3**j**, the angular dependence of a typical peak ($B^* = 0.7$ T) shows anisotropies $\frac{B^*(a)}{B^*(b)} \sim 9$ and $\frac{B^*(c)}{B^*(b)} \sim 7$, respectively, infers a less anisotropic dispersion than anisotropic ratios 13, 8 in a previous report[17].

The two observations drive us to focus on the specific energy dispersion in the magnetic field. As we know, the Landau levels of the Dirac band, in **ac** plane of ZrTe$_5$, are expressed as: $E_n = \sqrt{2nBv_xv_ye\hbar}$, where $e$ is the elemental charge. With fixed Fermi energy, we get a relation of $n \sim \frac{1}{B}$. However, the Zeeman splitting $\frac{\bar{g}}{2}\mu_B B$ is large (for simplicity, we here use averaged $g$-factor $\bar{g} \sim 15$[20,37]), results in a modified Landau level dispersion for the valence band: $E_{n(+)} = -\sqrt{2nBv_xv_ye\hbar} + \frac{\bar{g}}{2}\mu_B B$. Now, we get $n \sim \frac{C_0}{B} + C_1 * B$, shows that the Zeeman effect is crucial for the specific $n$-$B$ relation observed in this work. Due to the high Fermi velocity $v_F \sim 5 \times 10^5$ m/s[21], the energy scale of the kinetic part ($\sqrt{2nBv_xv_ye\hbar}$) is 18 meV*$\sqrt{nB}$, and the Zeeman splitting part ($\frac{\bar{g}}{2}\mu_B B$) is 0.6 meV*$B$, then the $C_1$ is usually very small and overcome by a parabolic band mixing effect for non-ideal Dirac dispersion[38]. However, the appearance of oscillations beyond the quantum limit indicates that the kinetic part should be heavily quenched in the magnetic field. Otherwise, oscillations can only appear at unreal magnetic fields[39]. Quenched kinetic energy, or equivalently flattening the band, in Dirac dispersion is parameterized by the enhanced mass and reduced Fermi velocity, which is already indicated by the reduced anisotropy $\frac{B^*(a)}{B^*(b)}$.

In the following, we will show that the kinetic energy of Dirac fermions in ZrTe$_5$ is heavily quenched in magnetic fields. Figure 4**a** shows the temperature-dependent oscillatory component $\Delta\rho_{xx}$ in sample S74. One prominent behavior is the high-field oscillations dampen quicker than the low-field ones. By extracting the amplitudes of peaks and valleys at characteristic fields, as shown in Fig. 4**b**, the temperature dependence of amplitudes $\Delta\rho_{xx}$ can be well fitted by the temperature-damping prefactor $\lambda = \frac{2\pi^2 k_B m_c}{\hbar e B} T$ of the Lifshitz–Kosevich (L-K) formula[40], where $k_B$ is the Boltzmann constant. As shown in Fig. 4**c**, the resulting cyclotron mass ($m_c$) get heavily enhanced (ratio~$10^2$) in magnetic fields. The temperature-damping prefactor $\lambda_D$ of the L-K formula for Dirac fermion[41] is written as $\lambda_D = \frac{2\pi^2 k_B |\mu|}{\hbar e B v_F^2} T$, where the $\mu$ is the chemical potential. Then, the cyclotron mass $m_c$ fitted by $\lambda$ effectively reflects the quantity $|\mu|/v_F^2$ in the Dirac system, this is consistent with the universal definition $m_c = \frac{\hbar^2}{2\pi} \frac{\partial A_k}{\partial E}$, where $A_k$ is the extremal area of orbital. By using the fixed carrier density ($p$) constrain in real systems, we come to $m_c = \hbar(6\pi^2 p)^{\frac{1}{3}}/v_F$, means enhanced $m_c$ corresponds to a reduction of $v_F$, supporting the formation of a topological flat band. Figure 4**d** shows the magnetic field dependence of $v_F$ (zero-field $v_F \sim 5\times10^5$ m/s), and the lowest value of $v_F$ is around $10^3$ m/s at ~20 T. These effects of cyclotron mass enhancement and $v_F$ reduction are similar to that in twisted bilayer graphene[4], while much stronger than those observed in the NLSM[42] and Kondo insulator[43].

Let's come to a picture based on the above experimental observations. As illustrated in Fig. 5**a**, the topmost Rashba-splitted band A flattens in a magnetic field, and the kinetic energy is heavily quenched. At the same time, the lower band is still dispersive with large kinetic energy and then goes up quickly in the magnetic field. The simplified band A, as a topological flat band, thus exhibits quenched kinetic energy and dominated Zeeman energy. Then, the Landau levels will bend and appear to FL. We plot the kinetic energy $E_{LL,kinetic} = -\sqrt{2nBv_xv_ye\hbar}$ (Fig. 5**b**) with fitted $m_c$, which quenches at small fields. As shown in Fig. 5**c**, we plot the total energy $E_n = -\sqrt{2nBv_xv_ye\hbar} + \frac{\bar{g}}{2}\mu_B B$ ($\bar{g} \sim 15$), and find the Landau levels cross the FL, distributing evenly in high magnetic fields. Therefore, the anomalous resistance oscillations observed beyond the quantum limit are consistent with forming a topological flat band in ZrTe$_5$.

In conclusion, our work demonstrates that the 3D Dirac material ZrTe$_5$ can be naturally transformed into a topological flat band system by symmetry breaking at low temperatures without invoking van der Waals heterostructure-based engineering. In this topological flat band, we observe anomalous resistance oscillations beyond the quantum limit, originating from the quenched kinetic energy of electrons on this flat band. Our work has two consequences: the first one is that the intrinsically formed topological flat band provides another route to create a topological flat band system beyond van der Waals heterostructures-based engineering; the second one is that the realization of a topological flat band in a 3D archetypical Dirac system opens the door for studying the specific exotic phenomena for topologically correlated effects uniquely existed in the dimensionality of three.

## Methods

### Crystal growth and characterizations

For single crystals of ZrTe$_5$ grown by the Te-flux method: Starting materials Zr (Alfa, 99.95%) and Te (99.9999%) were sealed in a quartz tube with a ratio of 1:300 and placed in a box furnace. Then, the materials were heated up to 900°C, followed by shaking the melt, and kept for two days. The temperature was then lowered to 660°C within 7 hours and cooled to 460°C in 200 hours. Single crystals of ZrTe$_5$ were isolated from Te flux by centrifuging at 460°C. Iterative temperature cycling was adopted to increase the size of ZrTe$_5$.

For single crystals of ZrTe$_5$ grown by CVT method: ZrTe$_5$ powder was synthesized by stoichiometric Zr (aladdin, 99.5%) and Te (99.999%). ZrTe$_5$ powder and iodine were sealed in a quartz tube and then placed in a two-zone furnace with a temperature gradient of 540 °C and 445 °C for one month.

The crystal structure and composition were checked by powder x-ray diffraction (XRD, Rigaku) and energy dispersive x-ray (Hitachi-SU5000).

### Transport measurements

Due to the reaction between silver paste and ZrTe$_5$, the usual bonding method with silver paste directly applied on the sample surface is not applicable here, which causes huge contact resistance and contaminates the electrical transport. To make good contacts, the surface was cleaned first by Ar plasma, 5 nm Ti/50 nm Au was then deposited on the *ac* surface with a homemade Hall bar mask. Then, silver paste was used to bond the Au wires for typical four-probe electrical contacts. The contact resistance is around several Ohms, and no sizable increase in contact resistance was observed over several months.

Electrical transport measurements above 2 K were carried out in a Quantum Design Physical Property Measurement System 9 T (PPMS-9 T) with a sample rotator. Measurements below 2 K were done in a top-loading dilution fridge (Oxford TLM, base temperature ~ 20 mK). High-field Measurements were carried out in a water-cooled magnet with steady fields up to 33 T in the Chinese High Magnetic Field Laboratory (CHMFL), Hefei. Lock-ins (SR830) were used to measure the resistance and the $2\omega$ resistance with a Keithley 6221 AC/DC current source.

**Torque magnetometry**

Magnetic torque measurements were carried out on a piezoresistive cantilever with a compensated Wheatstone bridge. The tiny sample cut from the pristine crystal was mounted on the tip of the cantilever. The resulted torque is roughly calculated by the relation: $\tau = \frac{4}{3}\frac{at^2}{\pi_L}\frac{\Delta V}{iR_s}$, where a is the leg width, t is the leg thickness, coefficient $\pi_L = 4.5 \times 10^{-10} m^2/N$, $\Delta V$ is the voltage drop on the Wheatstone resistance bridge, $i$ is the current fed into the bridge and $R_s$ is the resistance of sample leg.

**Acknowledgments** We thank Shuang Jia for providing cantilevers and supporting us on the initial construction of the torque setup. D. Xing thanks Jianghao Jin, Yuxin Yang and Tianping Ying for assisting on crystal growth, and Jianfei Xiao for helping with contact fabrication. C.-L. Zhang was supported by a start-up grant from the Institute of Physics, Chinese Academy of Sciences. J. Luo was supported by the National Science Foundation of China (Grants No. 12134018). J. Zhang was supported by the National Key R&D Program of the MOST of China (Grant No. 2022YFA1602602) and the Natural Science Foundation of China (Grants No. 12122411). A portion of this work was carried out at the Synergetic Extreme Condition User Facility (SECUF).

**Author contributions** C.-L.Z. conceived and supervised the project. D.X., S.P., B.T., Z.W., J.Z., J.L. and C.-L.Z. performed the electrical transport experiments; C.-L.Z. did the magnetic torque experiments. D.X. and C.-L.Z. grew the single-crystalline samples and characterized them by XRD and EDX; C.-L.Z. and D.X. analyzed the data. C.-L.Z. wrote the paper with input from all other authors.

**Additional information** Supplementary information is available in the online version of the paper. Reprints and permissions information are available online reprints. Correspondence and requests for materials should be addressed to C.-L.Z.

**Competing financial interests** The authors declare no competing financial interests.

**Data availability** The data supporting the plots within this paper and other findings are available from the corresponding authors upon reasonable request.

**Code availability** The computer codes used for the data analysis and numerical simulation are available from the corresponding authors upon request.

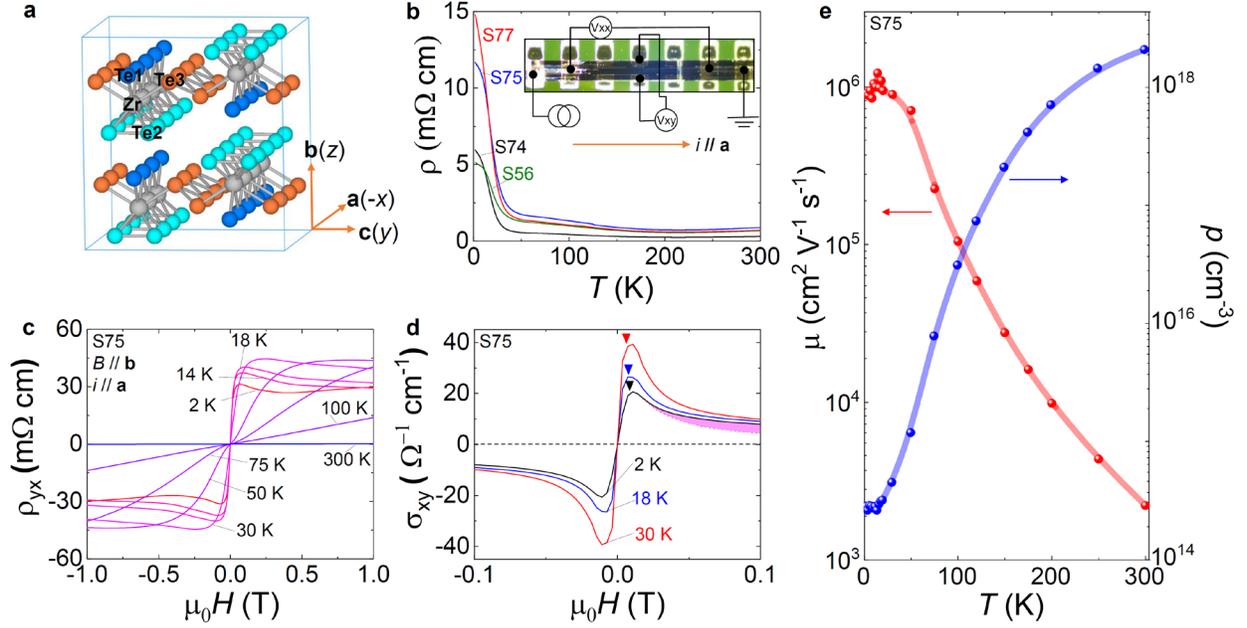

Fig. 1| **Electrical transport characterizations of flux-grown ZrTe$_5$. a,** Crystal structure of undistorted ZrTe$_5$. **b,** Temperature-dependent resistivity $\rho_{xx}$ of typical flux-grown ZrTe$_5$ samples measured in this work. **c,** Temperature-dependent Hall resistivity $\rho_{yx}$ of sample S75. Anomalous term develops when the temperature is low. **d,** Temperature-dependent Hall conductivity $\sigma_{xy}$ converted from $\rho_{yx}$. The anomalous term $\sigma_{xy}^A$, shadowed by the pink area, occurs after the Drude subtraction (dashed line). **e,** Temperature-dependent carrier density ($p$) and mobility ($\mu$) of sample S75.

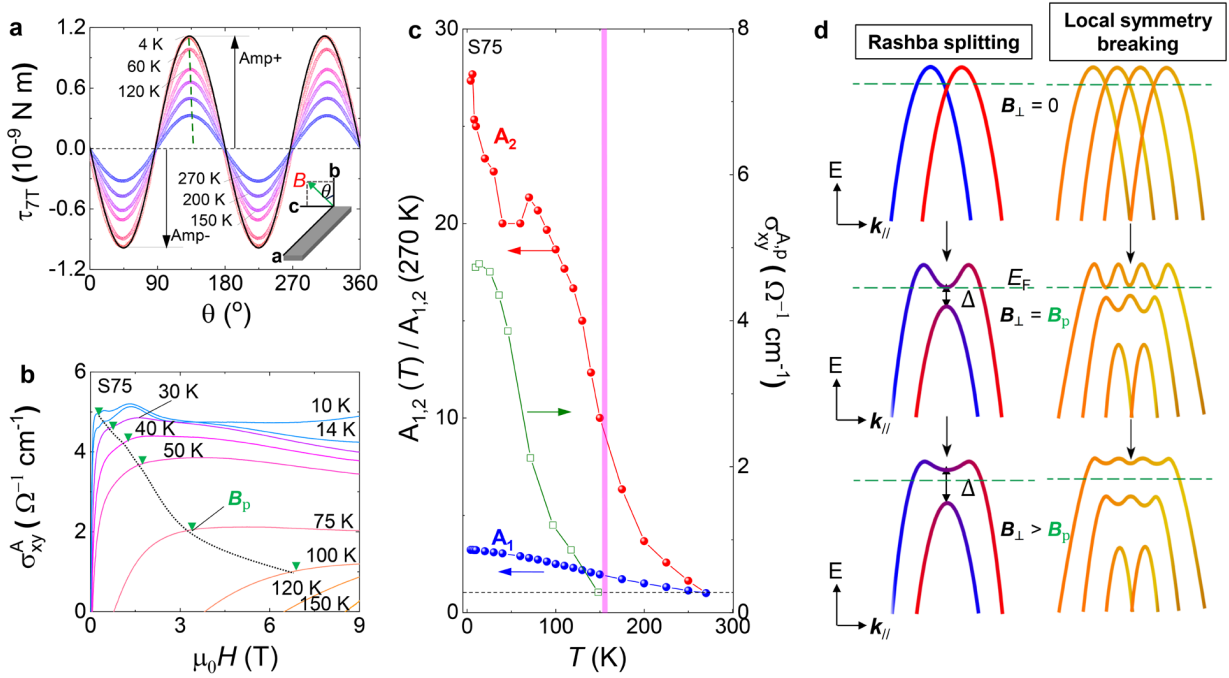

Fig. 2| **Polar-distortion-induced Rashba band splittings in ZrTe$_5$. a,** Angle-dependent magnetic torque at 7 T measured at different temperatures. The inset illustrates the experimental setup. The solid black line shows the fitting by the formula $\tau_a = \tau_{2\theta} + A_2 \sin^2\theta$. **b,** Temperature-dependent anomalous Hall conductivity $\sigma_{xy}^A$ obtained after subtracting the Drude component. **c,** Temperature-dependent $A_{1,2}(T)/A_{1,2}(270\,K)$ and $\sigma_{xy}^A$ in sample S75. The dashed vertical pink line indicates the concurrence of $\sigma_{xy}^A$ and polar distortion. **d,** Illustration of band modifications in the presence of polar distortion. The left-hand column shows the magnetic field-induced gap in the original Rashba bands, and finally, a local topological flat band is formed at the top of the band in a higher magnetic field. The right-hand column shows the flatness is strongly enhanced by the local symmetry breaking promoted by polar distortions.

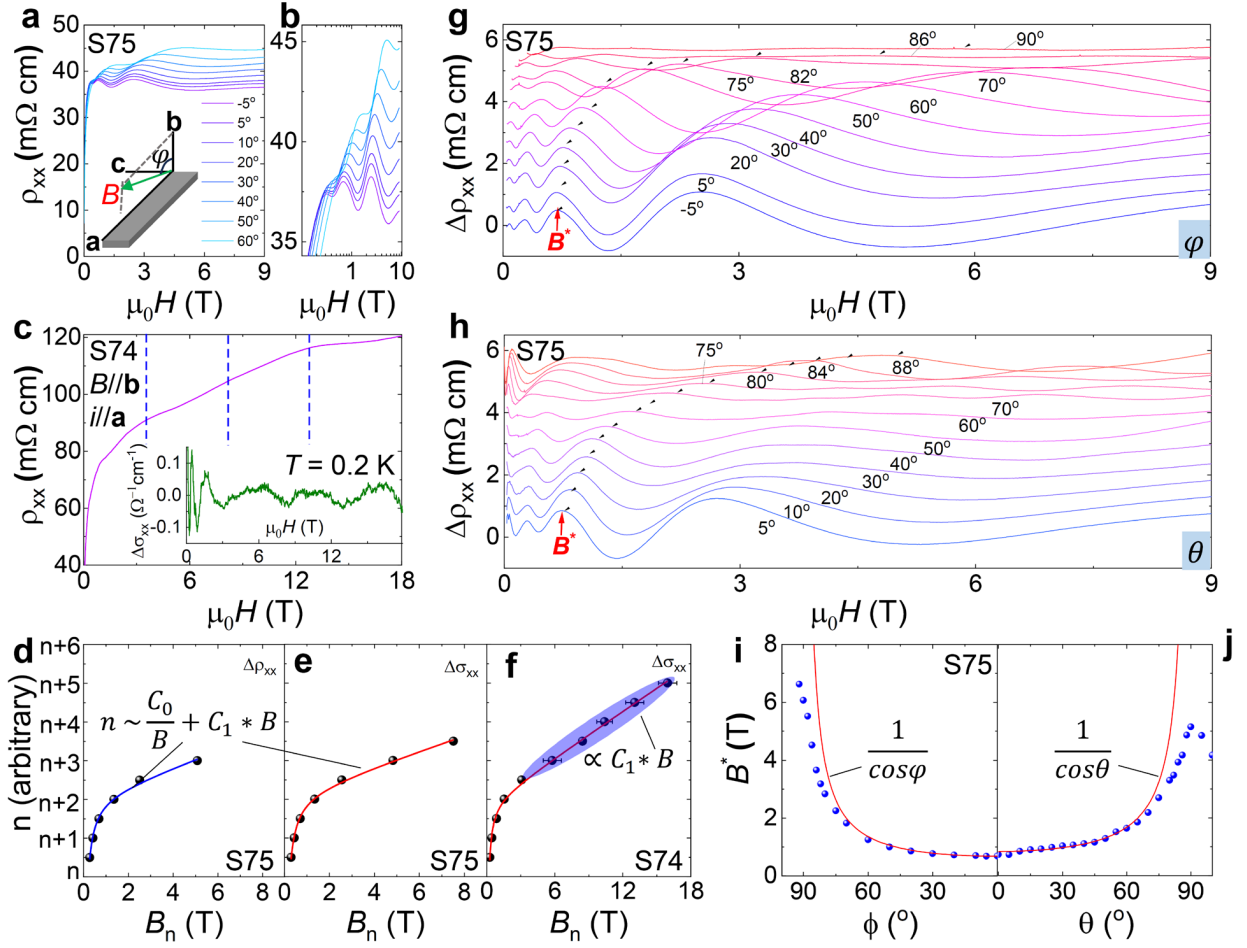

Fig. 3| **Anomalous resistance oscillations beyond the quantum limit in ZrTe$_5$. a&b,** Magnetoresistance measured at different angles with standard and logarithmic scales in sample S75, respectively. The inset shows the experimental setup for rotation. **c,** Magnetoresistance measured up to 18 T at 0.2 K in sample S74. Inset shows conductivity oscillations $\Delta\sigma_{xx}$ obtained from background subtraction. **d-f,** Landau fan diagrams with arbitrary integers based on oscillatory components $\Delta\sigma_{xx}$ and $\Delta\rho_{xx}$, respectively. Solid blue and red lines are the fittings by the relation $n \sim \frac{C_0}{B} + C_1 * B$. **g&h,** Oscillatory component $\Delta\rho_{xx}$ of tilted angles $\theta$ and $\phi$, respectively. $B^*$ denotes the characteristic peak ($B^* = 0.7$ T). **i&j,** The $\theta$ and $\phi$ dependence of characteristic peak $B^*$. Red lines are $1/\cos\theta, \phi$ relations.

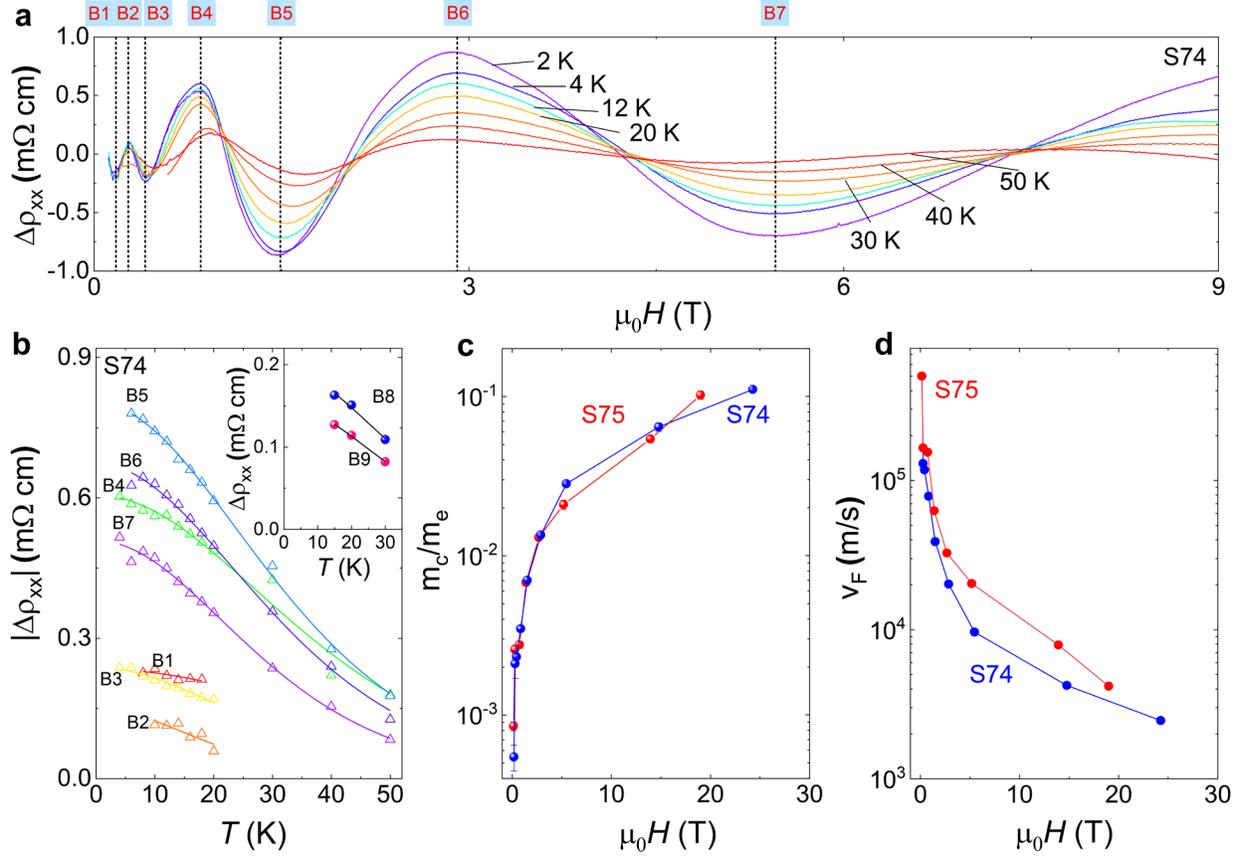

Fig. 4| **A topological flat band evidenced by field-induced mass enhancement in ZrTe$_5$.** **a,** Temperature-dependent oscillatory component $\Delta\rho_{xx}$ of sample S74. B1-B7 denote the seven characteristic peaks and valleys subjected for the L-K formula fitting. **a,** Temperature-dependent amplitudes of the component $\Delta\rho_{xx}$ fitted by the L-K formula. The inset shows the L-K formula fitting at fields higher than 9 T measured in a 33 T water-cooled magnet. **c&d,** Enhanced cyclotron mass ($m_c$) and reduced Fermi velocity ($v_F$) in the magnetic fields.

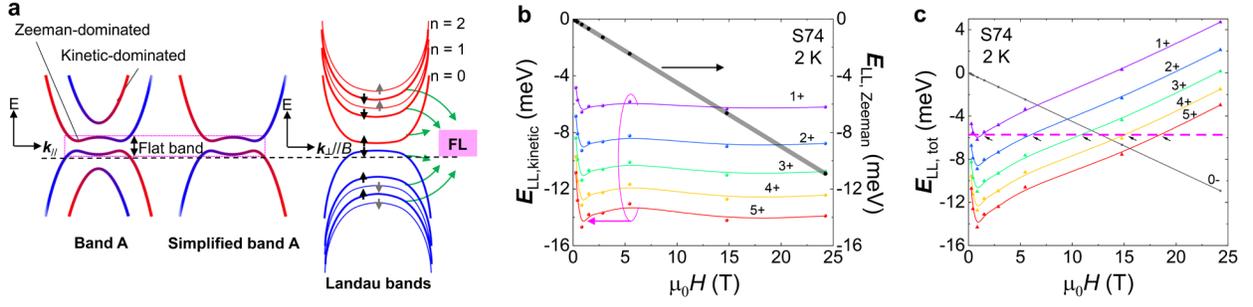

Fig. 5| **Quenched kinetic energy of Landau levels. a,** Formation of Landau levels on the topological flat band with dominated Zeeman effect. **b,** The kinetic energy, $E_{LL,kinetic} = -\sqrt{2nBv_xv_ye\hbar}$, of Landau levels obtained from experimental cyclotron mass, showing the quenched kinetic energy at high fields. **c,** The total energy of Landau levels, including the dominated Zeeman effect $\frac{\bar{g}}{2}\mu_B B$, exhibiting the reappearance of Landau levels across the FL.

# Supplementary Information

# Rashba-splitting-induced topological flat band detected by anomalous resistance oscillations beyond quantum limit in ZrTe$_5$

Dong Xing[1,2], Bingbing Tong[1], Senyang Pan[3], Zezhi Wang[1,2], Jianlin Luo[1,2], Jinglei Zhang[3], and Cheng-Long Zhang[1*]

[1]*Beijing National Laboratory for Condensed Matter Physics, Institute of Physics, Chinese Academy of Sciences, Beijing 100190, China*

[2]*School of Physical Sciences, University of Chinese Academy of Sciences, Beijing 100049, China.*

[3] *High Magnetic Field Laboratory, HFIPS, Chinese Academy of Sciences, Hefei 230031, China*

Corresponding   author: chenglong.zhang@iphy.ac.cn

# 1. Torque data in CVT samples and related symmetry analyses

We also performed the same torque measurement on a sample grown by the chemical vapor transfer (CVT) method. As shown in Fig. S1**b**, the angular torque at 7T displays a negligible asymmetric amplitude reflecting a small $A_2$ term. By adopting the same fitting process as outlined in the main text, as shown in Fig. S1**c**, the ratio $A_2(T)/A_2(250\ K)$ exhibits no enhancement at low temperatures and follows the ratio $A_1(T)/A_1(250\ K)$, which is distinct from the results of Fig. 2**c** in the main text and Fig. S1**a**.

We analyze possible crystal structures that flux-grown ZrTe$_5$ samples adopt in low temperatures. Angular torque is powerful for systems with orthogonal axes, like cubic, tetragonal, or orthorhombic structures. Our torque results show the measured results are incompatible with an orthorhombic structure. For a monoclinic structure with a C$_2$ operation, the $\chi$ tensor adopts a form:

$$\chi_{ij} = \begin{pmatrix} \chi_{xx} & \chi_{xy} & 0 \\ \chi_{xy} & \chi_{yy} & 0 \\ 0 & 0 & \chi_{zz} \end{pmatrix}$$

Here *x, y, z* are used, and the magnetic torque $\tau$ is:

$$\vec{\tau} = \frac{1}{2}\mu_0 V H^2 \begin{pmatrix} \sin 2\theta \cdot (\chi_{yy} - \chi_{zz}) \\ \sin 2\theta \cdot (-\chi_{xy}) \\ \sin^2\theta \cdot (2\chi_{xy}) \end{pmatrix}$$

For a triclinic structure with no additional symmetry operations, the $\chi$ tensor adopts a general form:

$$\chi_{ij}(C_1) = \begin{pmatrix} \chi_{xx} & \chi_{xy} & \chi_{xz} \\ \chi_{yx} & \chi_{yy} & \chi_{yz} \\ \chi_{zx} & \chi_{zy} & \chi_{zz} \end{pmatrix}$$

and the magnetic torque $\tau$ is:

$$\tau_x = \frac{1}{2}\mu_0 V H^2[(\chi_{yy} - \chi_{zz}) \cdot \sin 2\theta + 2\chi_{yz} \cdot \cos 2\theta]$$

$$\tau_y = \frac{1}{2}\mu_0 V H^2(-\chi_{xy} \cdot \sin 2\theta - 2\chi_{xz} \cdot \cos^2 \theta)$$

$$\tau_z = \frac{1}{2}\mu_0 V H^2(\chi_{xz} \cdot \sin 2\theta + 2\chi_{xy} \cdot \sin^2 \theta)$$

ZrTe$_5$ adopts a crystal structure with lower symmetry, like monoclinic or triclinic with non-orthogonal axes, complicating the angular magnetic torque measurements and analyses. Nevertheless, in the above analyses, we can see there indeed appears $\sin^2 \theta$ term when the magnetic field rotates in the *yz* (*bc*) plane, namely the A$_2$ term, in the $\chi_{ij}$ tensor of the two crystal structures with lower symmetries.

## 2. Nonreciprocal transport measured under negligible Joule heating

We performed the nonreciprocal transport measurements on sample S75 with magnetic fields along *b* and *c* directions, respectively. In the measurements process, we found that the magnetic-field-symmetric Seebeck effect contributes to the raw signal even with a current as low as 0.1mA. The Seebeck effect is subtracted in the process of field antisymmetrization. While increasing the excitation current, a magnetic-field-asymmetric signal arises at small fields, which superimposes on a linear background after antisymmetrization. The Nernst effect causes this anomaly at a small field due to an unevenly distributed thermal gradient caused by contact resistance Joule heating. We have to set an excitation current of *i* = 0.1 mA to minimize this effect. As we know, nonreciprocal resistance[1] adopts a form $R(I_0, B) \propto \gamma R_0 I_0 B$, where $\gamma$ is the coefficient that characterizes the strength of nonreciprocal resistance. We adopted $2\omega$ measurements[2,3], which produces $V_{2\omega} = \frac{1}{2}\gamma R_0 B I_0^2$, exhibiting a linear-in-*B* second-harmonic voltage. As shown in Fig. S2**a**, We now can recover this linear-in-*B* second-harmonic voltage and find that the nonreciprocal signal $V_{xx}^{2\omega}$ (B//*c*) is larger than $V_{xx}^{2\omega}$ (B//***b***). This means there is a polar component (***P***) along the out-of-plane ***b*** axis in our samples, unlike the negligible polarity along the ***b*** axis in ref[4].

## 3. Angle-dependent $\rho_{yx}$ for ***ba*** and ***bc*** planes

As shown in Fig. 3**a** and 3**b**, we measured angle-dependent Hall resistivity $\rho_{yx}$ in both ***ba*** and ***bc*** planes, respectively. There is no in-plane anomalous Hall (AHE) signal in our flux-grown samples. Our samples are different from those used in ref[5], but similar to those used in ref[6]. Our results are also consistent with the angular $\rho_{yx}$ reported in ref[6]. Therefore, the AHE appears when the magnetic field is along ***b*** axis, and suddenly disappears when the magnetic field is tilted along ***ac*** in-plane configuration, supporting the Rashba-splitting picture proposed in the main text.

## 4. Process of background subtraction for obtaining the oscillatory $\Delta\rho_{xx}$

As shown in Fig. S4, the magnetoresistance increases very quickly at small fields and tends to saturate at higher fields, which is found to be well fitted by exponential functions $c * e^{-(x-x_0)/t_1}$, which is a smooth function shown by the black fitting line.

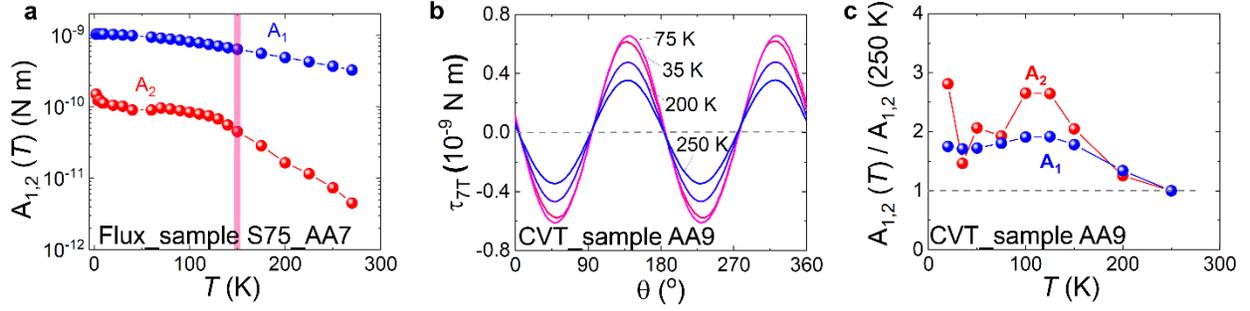

FIG. S1: **Raw data of Fig. 2c and magnetic torque measurements in CVT-grown sample. a**, Raw data of $A_1$ and $A_2$ for Fig. 2c in the main text, showing the $A_2$ term suddenly appears around T =150 K, while A1 is smooth and almost unchanged in the whole temperature range. **b**, Angle-dependent magnetic torque measured on a CVT-grown sample. **c**, Ratios of $A_{1,2}(T)/A_{1,2}(250\,K)$ for CVT-grown sample, showing no anomaly on $A_2$, which also follows the temperature dependence of $A_1$.

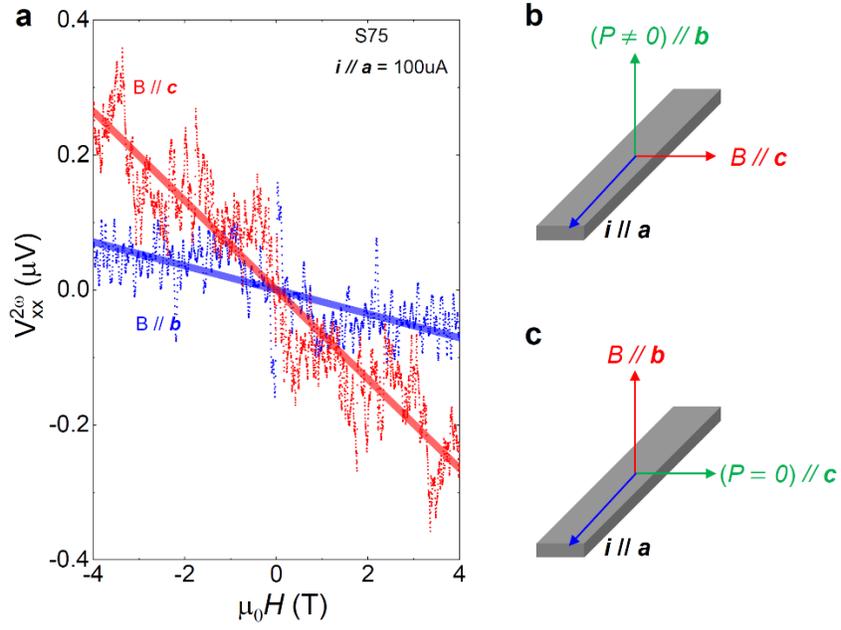

FIG. S2: **Nonreciprocal resistance measurement in sample S75. a**, Second harmonic voltage $V_{xx}^{2\omega}$ for B//*c* and B//*b*, respectvely. **b** & **c**, Appearance of linear-in-B $V_{xx}^{2\omega}$ for the magnetic field along *c* axis indicates there is a polarity component along *b* axis.

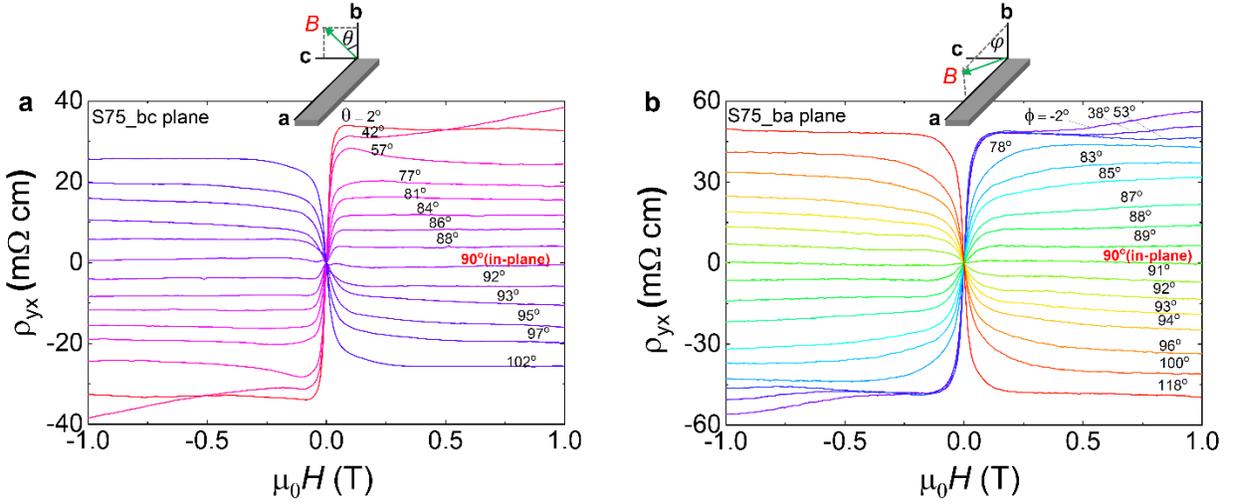

FIG. S3: **Angle dependence of $\rho_{yx}$ in *bc* and *ba* planes for sample S75. a**, $\rho_{yx}$ measured in *bc* plane with rotating angle $\theta$ (inset). $\theta = 90^o$ corresponds to the magnetic field along **c** axis. **b**, $\rho_{yx}$ measured in *ba* plane with rotating angle $\phi$ (inset). $\phi = 90^o$ corresponds to the magnetic field along **a** axis.

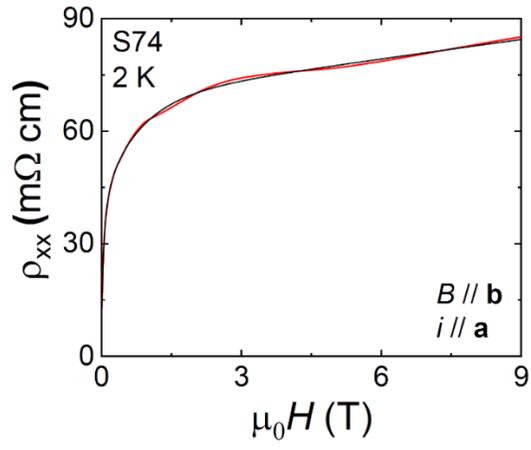

FIG. S4: **Background subtraction for obtaining the oscillatory $\Delta\rho_{xx}$.**

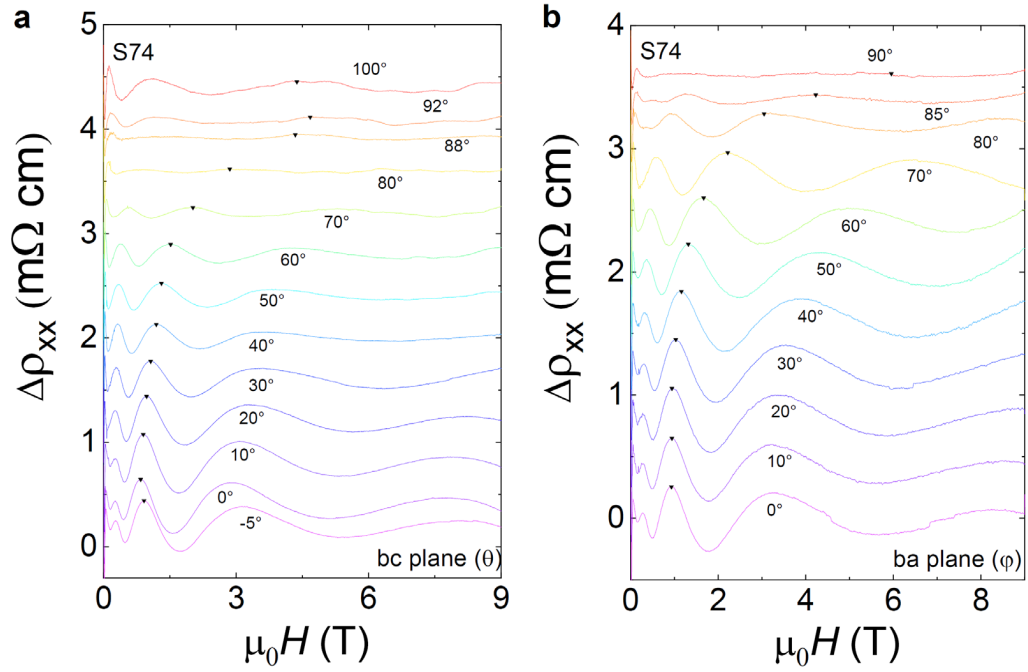

FIG. S5: **Angle dependence of $\Delta\rho_{xx}$ for sample S74 in *bc* and *ba* planes.**

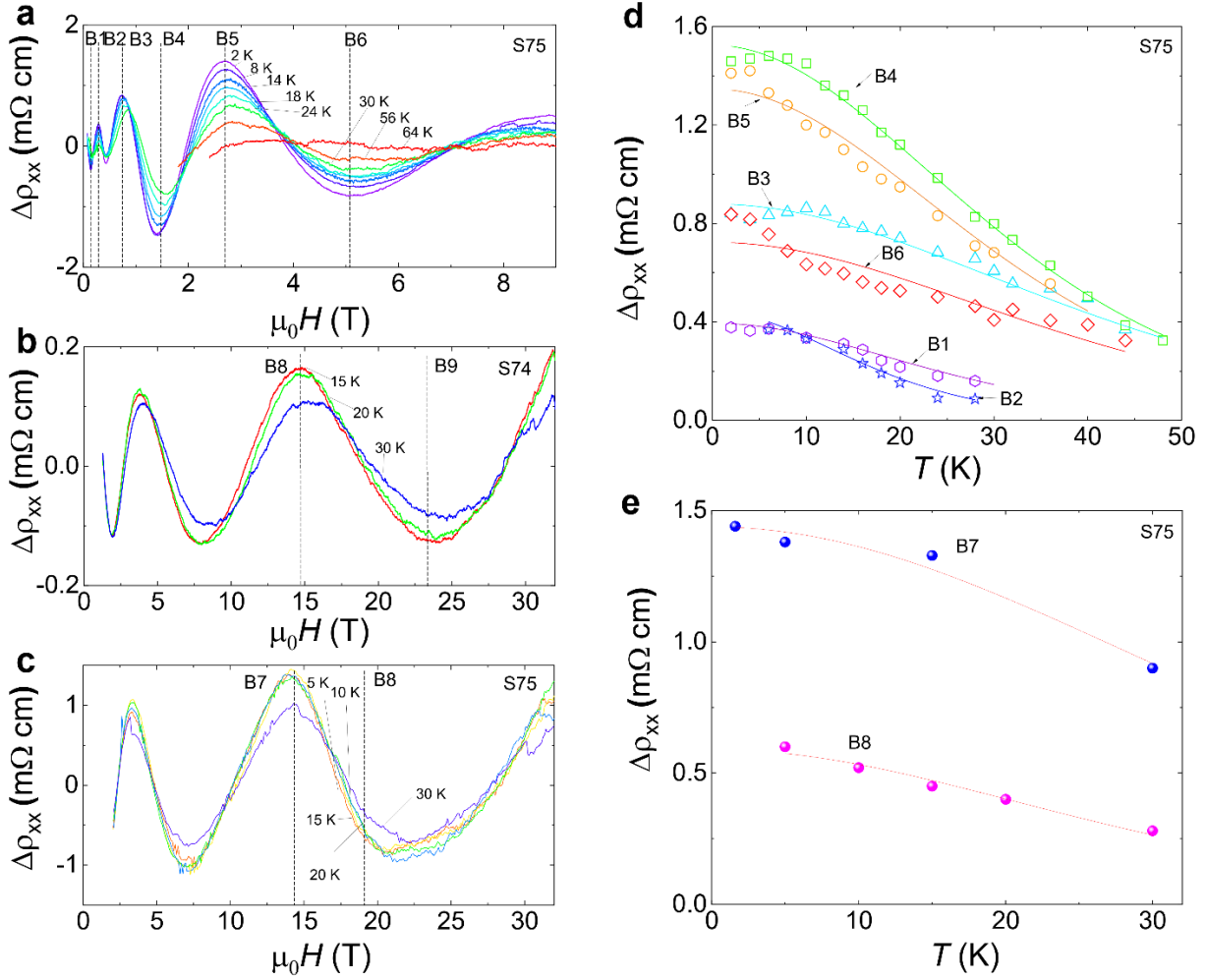

FIG. S6: **Additional data of temperature-dependent $\Delta\rho_{xx}$. a**, Temperature-dependent $\Delta\rho_{xx}$ of S75. **b** & **c**, Temperature-dependent $\Delta\rho_{xx}$ of samples S74 and S75 measured in high fields. **d** & **e**, The L-K formula fittings for sample S75